\documentclass[12pt]{iopart}

\usepackage{graphicx}
\begin{document}

\title{The conjunction fallacy and interference effects}

\author{Riccardo Franco
\footnote[3]{To whom correspondence should be addressed
riccardo.franco@polito.it}}
\address{Dipartimento di Fisica and U.d.R. I.N.F.M., Politecnico di Torino
C.so Duca degli Abruzzi 24, I-10129 Torino, Italia}

\date{\today}

\begin{abstract}
In the present article we consider the conjunction fallacy, a well known cognitive heuristic experimentally tested in
cognitive science, which occurs for intuitive judgments in situations of bounded rationality. We show that the quantum
formalism can be used to describe in a very simple way this fallacy in terms of interference effect. We evidence that
the quantum formalism leads quite naturally to violations of Bayes' rule when considering the estimated probability of 
the conjunction of two events. By defining the concept of maximal conjunction error, we find a good agreement with experimental results. Thus we suggest that in cognitive science the formalism of quantum mechanics can be used 
to describe a \textit{quantum regime}, the bounded-rationality regime, where the cognitive heuristics are valid.
\end{abstract}

\maketitle

%
\section{Introduction}
This article addresses two main directions of research: the investigation of how the quantum formalism is compatible
with Bayes' rule of classic probability theory, and the attempt to describe with the quantum formalism systems and
situations very different from the microscopic particles. A number of attempts have been done to apply the formalism of
quantum mechanics to domains of science different from the micro-world with applications to
economics \cite{economy}, operations research and management sciences \cite{management} and \cite{Rfranco_rat_ign_1},
psychology and cognition \cite{psychology} and \cite{aerts3}, game theory \cite{game}, and language and
artificial intelligence \cite{language}.
Quantum mechanics, for its counterintuitive
predictions, seems to provide a good formalism to describe puzzling effects of contextuality. In the present article,
we try to describe within the quantum formalism an important heuristic of cognitive science, the conjunction
fallacy \cite{Kahn-Tversky-conj-fall, osherson}. This heuristic is valid in regime of bounded rationality, which is characterized
by cognitive limitations of both knowledge and cognitive capacity. Bounded rationality \cite{bounded} is a central
theme in behavioral economics and psychology and it is concerned with the ways in which the actual decision-making
process influences agents' decisions. Previous attempts to describe features connected with this heuristic in terms
of quantum formalism have been done in \cite{Rfranco_conj1} and in \cite{bordley}.

This article is organized in order to be readable both from quantum physicists and from experts of cognitive science.
In section \ref{section:formalism and test} we introduce the basic notation of quantum mechanics, and we show in
\ref{section:repeated_test} that the quantum formalism describing two non-commuting observables leads to violations of
Bayes' rule. In section \ref{section:bounded rationality} we describe the answers to a question in bounded-rationality
regime in terms of vector state and density matrix of the quantum formalism. Finally,   in section
\ref{section:conjunction} we show how the quantum formalism can naturally describe the conjunction fallacy,
and in section \ref{section:new}   we try to
describe with the quantum formalism other fallacies
%

The main results of this article are: 1) tests on non-commuting observables lead to violations of Bayes' rule; 2) the
opinion-state of an agent for simple questions with only two possible answers can be represented, in
bounded-rationality regime, by a qubit state; 3) the different questions in bounded-rationality regime can be formally
written as operators acting on the qubit states; 4) the explicit answer of an agent to a question in regime of bounded
rationality can be described as a collapse of the opinion state onto an eigenvector of the corresponding operator; 5)
the estimated probability relevant to a question $A$, when analyzed in terms of the probability relevant to a second 
question $B$
(corresponding to a non-commuting operator) evidences the violation of Bayes' rule. The conjunction fallacy thus
results as a consequence of this general fact. 6) The predictions of the model are consistent with the experimental tests
on conjunction fallacy.

In conclusion, we present a very general  and abstract formalism
which seems to describe the heuristic of conjunction fallacy. We
think that a similar study can be done for other heuristics of
cognitive science (this will be presented in new papers). Thus
these heuristics could be simple applications of a general theory
describing the bounded-rationality regime, which probably will
lead to new interesting predictions. This could confirm the
hypothesis  that the processes of intuitive judgement could
involve mechanisms at a quantum level in the brain.
\section{Quantum basic formalism}\label{section:formalism and test}
We first introduce the standard bra-ket notation usually used in
quantum mechanics,  introduced by Dirac \cite{Dirac}, and then the
density matrix formalism. In particular, we focus our attention on
the concept of qubit. In the simplest situation, a quantum state
is defined by a ket $|s\rangle$, which is a vector in a complex
separable Hilbert space $H$. If the dimension of $H$ is 2, the
state describes a qubit, which is the unit of quantum information.
Any quantum system prepared identically to $|s\rangle$ is
described by the same ket $|s\rangle$.

In quantum mechanics, we call a measurable  quantity an
\textit{observable}, mathematically described by an operator, for
example $\widehat{A}$, with the important requirement that it is
hermitian: $\widehat{A} = \widehat{A}^{\dag}$, where
$\widehat{A}^{\dag}$ is the conjugate transpose. In the case of a
single qubit,  any observable $\widehat{A}$ has two real
eigenvalues $a_0$ and $a_1$ and two corresponding eigenvectors
$|a_0\rangle$ and $|a_1\rangle$. Another property of hermitian
operators is that its eigenvectors, if normalized, form an
orthonormal basis, that is
\begin{equation}\label{orthonormality}
\langle a_i|a_j\rangle = \delta(i,j)\,, \end{equation}
where $i,j=0,1$ and $\delta(i,j)$ is the Kroneker delta, equal to
1 if $i=j$ and null otherwise.
Given such a basis in the Hilbert space, we can write them in
components as
\begin{equation}\label{vector}
|a_0\rangle = \left (
  \begin{tabular}{c}
    1 \\
    0
    \end{tabular} \right ) \,,\,\,
|a_1\rangle = \left (
  \begin{tabular}{c}
    0 \\
    1
    \end{tabular} \right )\,\,,
\end{equation}
representing the quantum analogue to the two possible values 0 and
1 of a classical bit.
%
An important difference is that in the quantum case a state can be
in a linear superposition of 0 and 1, that is
\begin{equation}\label{superposition_A} |s\rangle=s_0 |a_0\rangle + s_1 |a_1\rangle\,,
\end{equation}
with $s_0$ and $s_1$ complex numbers. We also say that the  state
$|s\rangle$ is a superposition of the eigenstates $|a_i\rangle$. In the
vector representation generated by formula (\ref{vector}) in the chosen
basis, the ket $|s\rangle$ and its dual vector, the bra $\langle s|$, can
be written respectively as
\begin{equation}\label{ket_bra_superposition}
|s\rangle= \left (
  \begin{tabular}{c}
    $s_0$ \\
    $s_1$
    \end{tabular} \right )
\,,\,\,\, \langle s|= \left (
  \begin{tabular}{cc}
    $s_0^*$ & $s_1^*$
    \end{tabular} \right )\,.
\end{equation}
Another mathematical object, which is important in order to
describe probabilities, is the inner product, also called
\textit{braket}. In general, the inner product of two kets
$|s\rangle$ and $|s'\rangle$ can be written, in the basis of the
obsevable $\widehat{A}$, as $\langle
 s|s'\rangle=s_0 {s'_0}^*+s_1 {s'_1}^*$, where $s'_i$ are the
components of $|s'\rangle$ in the same basis. Thus the  inner
product of $|s\rangle$ and its dual vector is $\langle
s|s\rangle=|s_0|^2+|s_1|^2$, and it is equal to 1 if the vector is
normalized.
Finally, $|s_i|^2=|\langle a_i|s\rangle|^2$ is the probability
$P(a_i)$ that the measure on the observable $\widehat{A}$ has the
outcome $i=0,1$.
The state $|s\rangle$ is called a \textit{pure} state, and
describes a quantum state for which the preparation is complete: all
the information which can be theorically provided have been used.

%
In the most general case, a quantum state is described by the
density matrix $\widehat{\rho}$, which is an hermitian operator
acting on $H$.
The density matrix $\widehat{\rho}$ describes in general  a
\textit{mixed state}, that is a state for which the preparation is
not completely determined. For example, the state may be in the
preparation $|s_1\rangle$ with a probability $P_1$, and in the
preparation $|s_2\rangle$ with a probability $P_2$ (the two
vectors may  be not orthogonal). We also say that the the mixed
state is a (statistical) mixture of the two states (or of the two
preparations):
\begin{equation}\label{mixed}
\widehat{\rho}=\sum_{i}P_i |s_i\rangle \langle s_i|\,.
\end{equation}
In the particular case where there is only one $P_i=1$, we have
$\widehat{\rho}=|s\rangle \langle s|$, that is a pure state: thus the
density matrix of a pure state is a projector. The
opposite situation is when the eigenvalues  of the density matrix
are all equal. In the single-qubit example, they are both $1/2$,
and the resulting operator is the identity matrix acting on the
Hilbert space $H$:
\begin{equation}\label{maximally_mixed}
\widehat{\rho}= \frac{1}{2} \widehat{I}= \frac{1}{2}\left[
    \begin{tabular}{cc}
    1 & 0 \\
    0 & 1
    \end{tabular}\right]\,.
\end{equation}
The resulting state is called maximally mixed, and can be
considered as the situation where the actual knowledge of the
state is null.
\subsection{Probability and mean value}
The elements of the single-qubit density matrix $\widehat{\rho}$
can be expressed in the basis of $\widehat{A}$, with $i,j=0,1$, as
\begin{equation}\label{mixed_elements}
\rho_{i,j}=\langle a_i|\widehat{\rho}|a_j\rangle \,.
\end{equation}
where the diagonal elements $\rho_{i,i}$ represent the probability
$P(a_i)$ to measure a certain value $a_i$ of the relevant
observable. An equivalent expression of these probabilities can be
written in terms of the trace-matrix operation:
\begin{equation}\label{probabilities}
P(a_i)=Tr(\widehat{\rho} |a_i\rangle \langle a_i|)=\rho_{i,i}
\end{equation}
This formula is the most general expression of the probability to
measure a value $a_i$ of an observable, when operating on quantum
systems identically prepared in the state $\widehat{\rho}$.
%
The formalism of the density matrix helps us to write in the most
general form the mean value of an observable $\widehat{A}$ as
\begin{equation}\label{mean_value}
\langle \widehat{A} \rangle = Tr(\widehat{\rho} \widehat{A})\,,
\end{equation}
which becomes, by using formula (\ref{mixed_elements}) and the
basis vectors of $\widehat{A}$, $\langle \widehat{A}
\rangle=\sum_i a_i P(a_i)$.
\subsection{Collapse of the state vector}
One of the axioms \cite{Dirac} of quantum mechanics states that,
given an initial mixed  state $\widehat{\rho}$ and an observable
$\widehat{A}$ acting on a discrete Hilbert space, we can define
from the eigenvectors $\{|a_i\rangle\}$ of $\widehat{A}$ the
projection operators $\{|a_i\rangle\langle a_i|\}$, where for a
single qubit $i=0,1$. Thus if the measure of the observable
$\widehat{A}$ is the eigenvalue $a_i$, the state updates as
\begin{equation}\label{collapse}
\widehat{\rho}  \rightarrow |a_i\rangle\langle a_i|
\end{equation}
This general formula is valid for an orthonormal basis
$\{|a_i\rangle\}$, and defines the collapse of the initial state
onto the state vector $|a_i\rangle$. Of course, after the measurement,
a suitable unitary operation can change the resulting (collapsed) state.
At a first analysis, the collapse of the state may seem quite obvious. In fact,
the initial probability distribution $P(a_j)$ given
by equation (\ref{probabilities}) corresponding to the initial
state $\widehat{\rho}$, changes, by performing simple calculations, into
\begin{equation}\label{collapse_P1}
P(a_j) \rightarrow \delta(a_i,a_j)\,.
\end{equation}
This means that, after measuring a certain value $a_i$ of
$\widehat{A}$, the probability that the observable actually has
another value $a_j\neq a_i$ is null. This fact is valid also in
classic probability theory. Nevertheless, we will show in the next
subsection that the collapse leads to violation of classic laws of
probability theory when considering more than one observable.
\subsection{Non-commuting operators and Bayes' rule}\label{section:repeated_test}
In quantum mechanics the operators associated to the observables
may not commute: for example, given two operators $\widehat{A}$
and $\widehat{B}$, acting on the same Hilbert space $H$, the
ordered product $\widehat{A}\widehat{B}$ can be different from
$\widehat{B}\widehat{A}$: in this case the two operators do not
commute and we write $[\widehat{A},\widehat{B}]\neq 0$, where
$[\widehat{A},\widehat{B}]=\widehat{A}\widehat{B}-\widehat{B}\widehat{A}$
is the commutator of the two observables. The consequences of this
fact are very important, and lead to violation of Bayes' rule.
Let us consider for simplicity a single-quit system and the
eigenvectors $\{|a_i\rangle\}$ and $\{|b_i\rangle\}$ of
$\widehat{A}$ and $\widehat{B}$ respectively (with $i=0,1$). The
probability of measuring the value $a_i$ or $b_i$ for the
observables $\widehat{A}$ or $\widehat{B}$ respectively is given
by equation (\ref{probabilities}), that is:
\begin{equation}\label{prob_AB}
P(a_i)=Tr(\widehat{\rho} |a_i\rangle \langle a_i|); \,\,
P(b_i)=Tr(\widehat{\rho} |b_i\rangle \langle b_i|)\,.
\end{equation}
We now  consider the conditional probability $P(b_j|a_i)$, defined
as the probability to measure the observable $\widehat{B}$ with
value $b_j$, given the occurrence of a measurement of
$\widehat{A}$ with value $a_i$. In quantum mechanics, the
occurrence of a measurement of $\widehat{A}$ with result $a_i$
means that the actual state is $|a_i\rangle$, independently from
the initial state before the measurement. This is a consequence of
the quantum collapse, and leads to many differences from the
classic case. In quantum mechanics thus we have that
\begin{equation}\label{conditional_p}
P(b_j|a_i)=|\langle b_j|a_i\rangle|^2=P(a_i|b_j)\,.
\end{equation}
Conditional probabilities in quantum mechanics have other particular
features: from the resolution of identity
$|a_0\rangle \langle a_0|+|a_0\rangle \langle a_0|=I$ (where $I$ is the
identity matrix), we have that $P(b_1|a_0)+P(b_1|a_1)=1$ and
$P(b_0|a_0)+P(b_0|a_1)=1$, from which we have in general that $P(b_0|a_0)=P(b_1|a_1)$
and $P(b_1|a_0)=P(b_0|a_1)$. These properties are not valid in general in classic probability theory, and can be responsible of the inverse fallacy \cite{Rfranco_rat_ign_3}.
In classical probability theory two observables are statistically independent
when $P(b_j|a_i)=P(b_j)$: in the quantum formalism this is impossible in general.

Let us now consider the Bayes' rule, which defines the joint
probability to measure contemporarily the values $a_i$ and $b_j$
for observables $A$ and $B$ respectively:
\begin{equation}\label{joint_cond}
P(a_i)P(b_j|a_i)=P(b_j)P(a_i|b_j)=P(a_i,b_j)\,.
\end{equation}
This equation is very important in classical probability theory,
since it links the joint probabilities relevant to $\widehat{A}$
and $\widehat{B}$ to the conditional probabilities.
In quantum mechanics one can not measure contemporarily two
non-commuting operators. This means that the joint probability $P(a_i,b_j)$ can not be
univocally defined. What we can rigorously define is
\begin{equation}\label{joint_cond2}
P(a_i \rightarrow b_j)=P(a_i)P(b_j|a_i)\,,
\end{equation}
where $P(a_i \rightarrow b_j)$ is the probability to measure $a_i$
for the observable $\widehat{A}$ and then the answer $b_j$ for the
observable $\widehat{B}$. We call
$P(a_i \rightarrow b_j)$ the \textit{consecutive probability} to
measure $a_i$ and then $b_j$.
From a formal point of view, the
impossibility to define univocally a joint probability is evidenced from the fact that, by using the
equations (\ref{prob_AB}) and (\ref{conditional_p}), we have in
general that
\begin{equation}\label{joint_cond1}
P(a_i \rightarrow b_j) \neq P(b_j \rightarrow a_i)
\end{equation}
Equation (\ref{joint_cond1}) evidences that in quantum mechanics the Bayes' rule is violated. Many of the paradoxical
results of quantum mechanics  are due to this violation. In the present article, we will focus our attention on the
conjunction fallacy, which we will study in the next sections.
%
%
\section{Bounded rationality and Hilbert spaces }\label{section:bounded rationality}
The bounded rationality  \cite{bounded} is a property of an agent
(a person which makes decisions) that behaves in a manner that is
nearly optimal with respect to its goals and resources. In
general, an agent acts in bounded-rationality regime when there is
a limited time in which to make decisions, or when he is also
limited by schemas and other decisional limitations. As a result,
the decisions are not fully thought through and they are rational
only  within limits such as time and cognitive capability. There
are two major causes of bounded rationality, the limitations of
the human mind, and the structure within which the mind operates.
This impacts decision models that assume us to be fully rational:
for example when calculating expected utility, it happens that
people do not make the best choices. Since the effects of bounded
rationality are counterintuitive and may violate the classical
probability theory (and the Bayes' rule), we will often speak of
\textit{bounded-rationality regime} as the set of situations where
the bounded rationality is an actual property.

We will show that some typical behaviors of the bounded-rationality regime can be described in a very effective way by
the quantum formalism. We will study the opinion state of agents having the same
initial information in terms of estimated probabilities. In particular, we will assume that the opinion state of an 
agent can be represented as a qubit
state, that is in terms of a density matrix $\widehat{\rho}$ or, in simple cases,  of a ket $|s\rangle$ in a Hilbert
space $H$ of dimension 2.
As in quantum mechanics experiments, it  is important to define carefully the preparation of the opinion state. Any
previous information given to an agent before performing a test can be considered as the preparation of the opinion
state. When we repeat a test on more agents, it is important that their opinion state is (at least in  theory)
\textit{identically prepared}. We note here that it is not easy to prepare the opinion state of a number of people in
an identical state. Nevertheless, the quantum formalism can help us with the concept of mixed state.

The basic test in the context of bounded rationality is a
question. We consider a question $A$ for which the possible
answers can only be 0 or 1 (false or true), and we associate it to
an operator $\widehat{A}$ acting on the Hilbert space $H$. Like in
quantum mechanics, the question $A$ is an observable, in the sense
that we can observe an answer: thus, when speaking of questions,
we will consider directly the associated operator $\widehat{A}$.
The answers 0 and 1 are associated to the eigenvalues $a_0=0$ and
$a_1=1$ of $\widehat{A}$, while the eigenvectors $|a_0\rangle$ and
$|a_1\rangle$ correspond to the opinion states relevant to the
answers 0 and 1 respectively:
\begin{equation}\label{eig1}
\widehat{A}|a_0\rangle=0|a_0\rangle=0;\,\,\,\widehat{A}|a_1\rangle=|a_1\rangle\,\,.
\end{equation}
The eigenvectors $|a_0\rangle$ and $|a_1\rangle$ have a very
precise meaning: if the opinion state of an agent can be described
for example by $|a_1\rangle$, this means that the answer to the
question $\widehat{A}$ is 1 with certainty. If we repeat the same
question to many agents in the same opinion state (thus
identically prepared), each agent will give the same answer 1. If
instead the opinion state about the question is definitely 0, then
we have the eigenvector $|a_0\rangle$.
Any observable can be written in the basis of its eigenvectors as
$\widehat{A}=\sum_i a_i |a_i\rangle\langle a_i|$.
A superposition of the opinion states $|a_i\rangle$ about question
$\widehat{A}$ is, like in equation (\ref{superposition_A}),
$|s\rangle=s_0 |a_0\rangle + s_1 |a_1\rangle$, where
$|s_i|^2=|\langle a_i|s\rangle|^2$ is the probability $P(a_i)$
that the agent gives an answer $i=0,1$ to the question
$\widehat{A}$.
In the most general case, the opinion state can be represented as
a density matrix $\widehat{\rho}$. The probability that the answer
to the question $\widehat{A}$ is $a_i$ is given by equation
(\ref{probabilities}): $P(a_i)=Tr(\rho |a_i\rangle \langle a_i|)$.

We note that the formalism introduced is the same used to describe
questions in regime of rational ignorance
\cite{Rfranco_rat_ign_1}, where people choose to remain uninformed
about a question $\widehat{A}$. In fact, the bounded rationality
can be considered as a more general situation than the rational ignorance,
where the question is preceded by some additional information.
\section{The conjunction fallacy}\label{section:conjunction}
The \textit{conjunction fallacy}  is a well known cognitive heuristic which occurs in bounded rationality when some
specific conditions are assumed to be more probable than the general ones. More precisely, many people tend to ascribe
higher probabilities to the conjunction of two events than to one of the single events. The most often-cited example of
this fallacy found by Tversky and Kahneman \cite{Kahn-Tversky-conj-fall} is the case of Linda, which
we will consider carefully in the present article. The test is preceded by a brief personality sketch:
\textit{Linda was a philosophy major. She is bright and
concerned with issues of discrimination and social justice.}
After this preparation, the agents are asked if they judge more probable that Linda is a bank teller or
a feminist bank teller. The surprising result is that subjects judge more probable
that Linda is a feminist bank teller than a bank teller, even if this is in contrast with classical
probability theory.

According to Tversky and Kahneman, the explanation of this fallacy is that subjects assess
Linda to be more typical of feminist bank teller than of bank teller, and that this
assessment influences their probability judgments.
As Kahneman and Tversky have repeatedly pointed out, however, the intuition
of typicality differs structurally from the mathematical properties of a classic probability measure.
There is no fallacy in considering Linda more typical of
a feminist bank teller than of bank teller, there is only a conjunction effect.
The conjunction fallacy instead lies in equating the probability order
to the typicality order.
%
A more thorough test of the similarity hypothesis has been described in \cite{osherson}: the stimuli
were 14 brief personality sketches, one of which was the Linda sketch and
others of which were new. For each sketch, they constructed an incompatible conjunction, $A$-and-$B$,
and a compatible conjunction, $A$-and-$C$, such that $A$ was the same representative outcome for both
conjunctions (e.g., Linda is a feminist), $B$ was an unrepresentative outcome (e.g., Linda is a bank
teller), and $C$ was a representative outcome (e.g., Linda is a teacher). In terms of the Linda
example, $A$-and-$B$ would be "Linda is a feminist bank teller" and $A$-and-$C$ would be "Linda is a feminist
teacher." One sample of subjects rated the typicality of the propositions, $B$/$A$-and-$B$, $C$/$A$-and-$C$, and
a different sample of subjects rated the probability of these same propositions.
The result is that exists a positive correlation between the conjunction effect (involving the typicality)
and the conjunction fallacy (involving the probability).

%
The main competitors to the representativeness explanation of
conjunction errors are various probability combination models. The
simplest weighted averaging model asserts that $P(A,B) = \frac{v_1
P(A)+ v_2 P(B}{v_1+v_2}$, where $v_1$ and $v_2$ are positive
weights (Wyer, 1976). A modified version of this model uses belief
strengths, and not probabilities. However, the main feature of
this model is that it leads to $P(A)>P(A,B)>P(B)$. These
combination models have been modified in order to avoid some
inconsistencies of the model with the experiments.
Yates and Carlson (\cite{Yates}) proposed an alternative theory of
conjunction errors according to which these errors result from the application of improper mental
rules when combining the probabilities of individual propositions.
The results of Table III  show that conjunction errors may occur also when
events $A$ and $B$ are semantically unrelated: it is difficult to see how the judged probability of the conjunction
could be based on the representativeness, i.e., the typicality, of the conceptual combination.
Thus they proposed a signed summation model where the belief strength of
a conjunction is a sum of the strengths of its components $S(A,B)=S(A)+S(B)$,
(the belief strength scale $S$ is allowed to take on either positive or negative values).
This model seem to be the most similar to our quantum explanation.

In general, we consider two dichotomous questions $A$ and $B$, with
possible answers $a_0, a_1$ and $b_0, b_1$. The typical experiments of
\cite{Kahn-Tversky-conj-fall} and \cite{osherson} consist in  a
preparation of the opinion state, which provides some information to the
agent, and the following question: what is more probable or frequent
between  $a_1$ and $a_1$-and-$b_1$? The agents manifest in all
these experiments  a strict preference for the answer $a_1$-and-$b_1$:
this evidences the conjunction fallacy, since the Bayes' rule
(\ref{joint_cond}) entails that
\begin{equation}\label{Conj_fallacy_classic}
P(b_1)=P(a_0)P(b_1|a_0)+P(a_1)P(b_1|a_1)\geq P(a_1)P(b_1|a_1) \,.
\end{equation}
In other words, the conjunction of two events $a_1$ and $b_1$ is always less probable than of one of two events.
Nonetheless, the agents often consider more likely $a_1$-and-$b_1$ than $a_1$.

We now introduce the quantum formalism in order to show that the
conjunction fallacy can be described and interpreted in such a formalism.
In particular, we consider two operators $\widehat{A}$ and $\widehat{B}$,
associated respectively to the questions $A$ and $B$. Since $A$ and $B$
are dichotomous questions, we can  describe the opinion state of agents as
vectors in a two-dimensional Hilbert space. Both the eigenvectors of
$\widehat{A}$ and $\widehat{B}$, defined by equation (\ref{eig1}), form
two orthonormal bases of the Hilbert space $H$. Thus we can express the
eigenvectors of $\widehat{A}$ in the basis of the eigenvectors of
$\widehat{B}$, obtaining the general equations:
\begin{eqnarray}\label{A-B-basis}
|a_0\rangle=e^{i\xi}\sqrt{P(b_1|a_1)}|b_0\rangle+e^{i\phi}\sqrt{P(b_1|a_0)}|b_1\rangle\\\nonumber
|a_1\rangle=-e^{-i\phi}\sqrt{P(b_1|a_0)}|b_0\rangle+e^{-i\xi}\sqrt{P(b_1|a_1)}|b_1\rangle\,\,,
\end{eqnarray}
and vice-versa
\begin{eqnarray} \label{B-A-basis}
|b_0\rangle=e^{-i\xi}\sqrt{P(b_1|a_1)}|a_0\rangle-e^{i\phi}\sqrt{P(b_1|a_0)}|a_1\rangle\\\nonumber
|b_1\rangle=e^{-i\phi}\sqrt{P(b_1|a_0)}|a_0\rangle+e^{i\xi}\sqrt{P(b_1|a_1)}|a_1\rangle\,\,.
\end{eqnarray}
The transformations above are a change of basis, which can be described in
terms  of a special unitary operator $\widehat{U}$ (which preserves the
normalization) such that $\sum_{ij}U_{ij}|a_i\rangle=|b_j\rangle$. The
transformation is defined uniquely by the three independent parameters
$P(b_1|a_1)$, $\phi$ and $\xi$: in fact the other parameter
$P(b_1|a_0)=1-P(b_1|a_1)$ is not independent, as noted in section (\ref{section:repeated_test}).
The parameters $\phi$ and $\xi$ are useful when considering pure states and are
important in  interference effects.

Let us now consider as the initial state the following superposition
\begin{equation}\label{S-A}
|s\rangle=\sqrt{P(a_0)}|a_0\rangle + e^{i\phi_a}\sqrt{P(a_1)}|a_1\rangle
\end{equation}
where $\phi_a$ represents the phase relevant to the superposition of
the vectors $|a_i\rangle$. This pure state describes correctly
the probabilities for $\widehat{A}$, $P(a_i)=|\langle s|a_i\rangle|^2$. By using equation
(\ref{A-B-basis}), we can express this state in the basis of
$\widehat{B}$, obtaining
\begin{eqnarray}\label{S-b}
&|s\rangle=\\\nonumber
&[\sqrt{P(a_0)} \sqrt{P(b_1|a_1)}e^{i\xi}-\sqrt{P(a_1)}
\sqrt{P(b_1|a_0)}e^{i(\phi_a-\phi)}]|b_0\rangle +\\\nonumber
&[\sqrt{P(a_0)}
\sqrt{P(b_1|a_0)}e^{i\phi}+ \sqrt{P(a_1)} \sqrt{P(b_1|a_1)}e^{i(\phi_a-\xi)}]|b_1\rangle.
\end{eqnarray}
We now consider the probabilities $P(a_1)$, $P(b_1)$ and the  conditional
probability $P(a_1|b_1)$: from equation (\ref{S-b}), we have that $P(b_1)$
is $|\sqrt{P(a_0)}\sqrt{P(b_1|a_0)}e^{i\phi}+
\sqrt{P(a_1)}\sqrt{P(b_1|a_1)}e^{i(\phi_a-\xi)}|^2$, obtaining
\begin{eqnarray}\label{interference}
&P(b_1)=P(a_0)P(b_1|a_0)+P(a_1)P(b_1|a_1)+\\\nonumber
&2\sqrt{P(a_0)P(a_1)P(b_1|a_0)P(b_1|a_1)}cos(\phi+\xi-\phi_a)
\end{eqnarray}
The presence of the last term, known as the interference term
$I(s,A)$ can produce conjunction fallacy effects: in fact, if we
impose that $P(a_0)P(b_1|a_0)+I(s,A)<0$, we have
$P(b_1)<P(a_1)P(b_1|a_1)$. Thus the sign and the weight of the interference term
can determine the conjunction fallacy, while the parameter $\phi+\xi-\phi_a$
can give to this effect more or less strength. A positive
interference term enhances the prevalence of $P(b_1)$ on
$P(a_1,b_1)$, which can be considered a \textit{reverse
conjunction fallacy}.

It must be noted that the quantum explanation of the conjunction
fallacy evidences that the interference term can make $P(b_1)$
lower than $P(a_1)P(b_1|a_1)=P(a_1 \rightarrow b_1)$.  On the
contrary, $P(b_1)$ is always greater than $P(b_1)P(a_1|b_1)=P(b_1
\rightarrow a_1)$. Thus we add the hypothesis that, when agents
have to judge the probabilities of $b_1$ and $a_1$-and-$b_1$, they
first consider the element that differences the two situations
($a_1$) and then the \textit{composite} situation. This hypothesis
is able to explain why the agents compare in their judgements
$P(b_1)$ and $P(a_1 \rightarrow b_1)$.

The conjunction fallacy  can appear also when comparing $a_1$ and $a_1$-and-$b_1$, if we write
the same initial state in the basis of $\widehat{B}$
\begin{equation}\label{S-BB}
|s\rangle=\sqrt{P(b_0})|b_0\rangle + e^{i\phi_b}\sqrt{P(b_1)}|b_1\rangle
\end{equation}
The probability $P(a_1)$ can be written, with similar calculations, as
\begin{eqnarray}\label{interference_b}
&P(a_1)=P(b_0)P(a_1|b_0)+P(b_1)P(a_1|b_1)-\\\nonumber
&2\sqrt{P(b_0)P(b_1)P(a_1|b_0)P(a_1|b_1)}cos(\phi-\xi-\phi_b)
\end{eqnarray}
evidencing once again an interference term $I(s,B)$. If we want
the presence of conjunction fallacy $P(a_1)<P(b_1)P(a_1|b_1)$, we
impose $P(b_0)P(a_1|b_0)+I(s,B)<0$.
We thus define the conjunction error as
\begin{eqnarray}\label{conjerr1}
&\Delta(A,B,s)=P(a_0)P(b_1|a_0)+\\\nonumber
&2\sqrt{P(a_0)P(a_1)P(b_1|a_0)P(b_1|a_1)}cos(\phi+\xi-\phi_a)<0   \\
\label{conjerr2}
&\Delta(B,A,s)=P(b_0)P(a_1|b_0)-\\\nonumber
&2\sqrt{P(b_0)P(b_1)P(a_1|b_0)P(a_1|b_1)}cos(\phi-\xi-\phi_b)<0
\end{eqnarray}
If the both the conditions
(\ref{conjerr1}) and (\ref{conjerr2}) are verified, we have a double conjunction fallacy.
\section{Experimental results}\label{section:results}
The model we have developed, once we know the independent
parameters $P(a_1)$, $P(a_1|b_1)$, $\phi+\xi-\phi_a$, allows to predict with
certainty if there is or not a conjunction fallacy.
However the experimental results evidence that not all the agents
manifest a conjunction fallacy, which can be associated to a
probability. This can be explained with the  fact that in general
we have an incomplete preparation of the opinion state, which
determines a mixed state: we could have some agents into a state
$|s_1\rangle$ with probability $P_1$ and some other agents into
the state $|s_2\rangle$ with probability $P_2$, etc... The
resulting opinion state is in general a mixed state
$\rho=\sum_{i}P_i|s_i\rangle$ where each $|s_i\rangle$ can have
different $P(a_1)$ and $\phi_a$. We thus have different types of
interference terms, determining in an independent way the
conjunction fallacy with probabilities $P_i$ respectively.

We define the mean conjunction error $\Delta_{mean}(A,B)=\sum_i \Delta(A,B,s_i)P_i$:
it could be null, even if some agents may manifest the conjunction fallacy, meaning that
the perceived compatibility of two events is not similar for the agents in the experiment.

We can decompose the density matrix into
$\rho=\sum_{i}P'_i|s'_i\rangle \langle s'_i| + \sum_{j}P''_j|s''_j\rangle \langle s''_j|$
where the states $|s'_i\rangle$ manifest conjunction fallacy, while the others
do not manifest it. Thus the probability that an agent will exhibit a conjunction fallacy,
being prepared in the state $\rho$, is $\sum_i P'_i$.
In the next subsections, we will investigate the mean conjunction errors and the probability of conjunction
errors.
In particular, we consider the probabilities $P(a_1)/P(b_1)$ with the following couples: HH, HL, LH, LL, where
H means that the corresponding probability is high ($>0.5$), while  L is low ($<0.5$).
For example the couple HH describes a situation where both $P(a_1)$ and $P(b_1)$ are $>0.5$.

\subsection{Mean conjunction error}
One of the main experimental results in the tests of the conjunction
fallacy  is the mean conjunction error $P(a_1,b_1)-P(b_1)$: in \cite{osherson}
it is found that there are greater mean conjunction errors in situations HL than in HH.
In particular, the maximal error reported in \cite{osherson} (table I) for HL is 0.27, while for HH is 0.09.
These experimental results of \cite{osherson} can be explained in the following way:
we define the condition of \textit{maximal conjunction error} compatible
with the parameters  $P(a_1)$ and $P(b_1|a_1)$ the values of
$\phi+\xi-\phi_a$ such that $cos(\phi+\xi-\phi_a)=-1$. By imposing the maximal conjunction
error condition we obtain the equation
\begin{equation}\label{condition1}
\Delta_{max}(A,B)=P(a_0)P(b_1|a_0)-2\sqrt{P(a_0)P(a_1)P(b_1|a_0)P(b_1|a_1)}<0
\end{equation}
where $\Delta_{max}(A,B)$ is the maximal conjunction error.
This difference is a measure of the strength of the maximal conjunction
error compatible with the parameters $P(a_1)$ and $P(b_1|a_1)$, which may
be lowered by different relative phase factor.
\begin{figure}[h]\label{fig1}
\centering
\includegraphics[width=11.5cm]{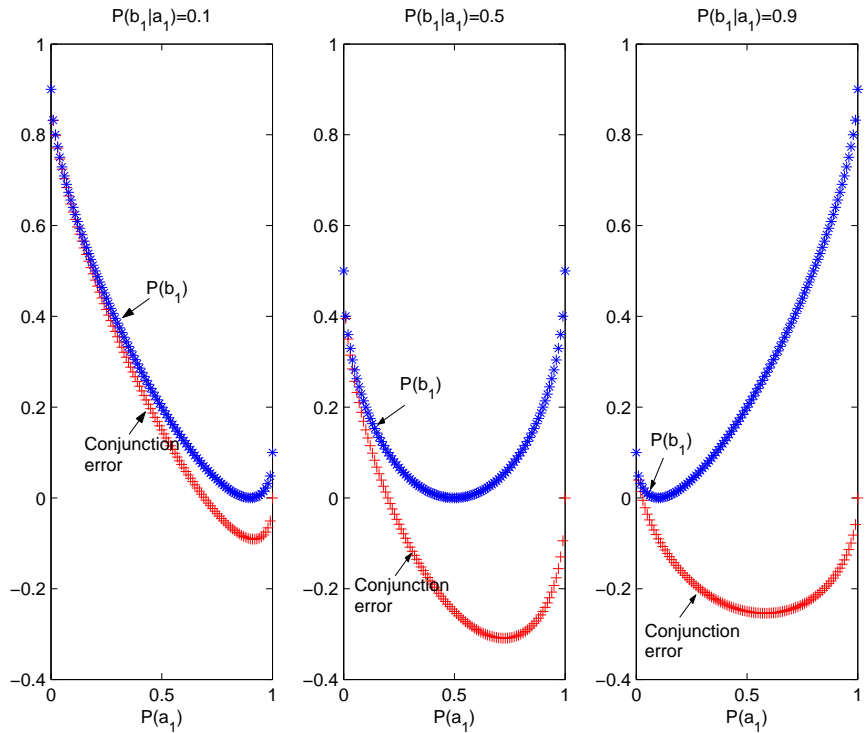}
\caption{Maximal conjunction errors for three different values of
$P(b_1|a_1)$}
\end{figure}
In figure 1 we show the maximal conjunction error for three
different values of $P(b_1|a_1)$ (0.1, 0.5 and 0.9). We note that
high/low values of $P(b_1|a_1)$ entail a
correlation/anticorrelation between $P(a_1)$ and $P(b_1)$
respectively.

Moreover, condition (\ref{condition1}) and figure 1 entail that
the combination LH does not manifest conjunction errors (the
maximal conjunction error is positive in the left part of the left
panel in figure 1), while the other couples allow for conjunction
errors. HH and LL couples produce lower maximal errors (0.17 and
0.23 respectively) than HL couples (0.3).

Now we consider the effect of the phase factor $cos(\phi+\xi-\phi_a)$: the conjunction error can be lowered
(negative factor), or can it be deleted (null or positive factor).
Thus the factor $cos(\phi+\xi-\phi_a)$ can be interpreted as a psychologic parameter of the
the opinion state of agents which describes how two events are considered antithetical.
For example, if two events are considered antithetical, we have a strongly negative interference term.
Our description could be connected to the concept of incompatibility of two events, introduced by
\cite{osherson}: two events are incompatible when we have a situation HL and
the phase factor is negative. It is important to distinguish
this kind of incompatibility from the incompatibility of quantum observables (that is
the non-commutativity).
It is important to note that the phase factor, because of its psychologic nature may be extremely
variable even for similar preparations.

Thus the results of \cite{osherson} can be explained by noting that for HL couples the predicted
maximal conjunction error (0.3) can be slightly lowered by the phase factor in order to be similar to the results of
\cite{osherson} (0.27). Analogously, for HH situations the maximal conjunction error predicted by the model
(0.17) is lowered by the phase factor in order to be similar to the experimental value 0.9.
We conclude that the phase factor is more similar to -1 in the HL case, confirming that the two events are
considered antithetically.

The case LH, according to our model, does not produce conjunction error, even without the effect of the phase factor.
This seems to be in accordance with experimental results, even if there is no general agreement about this fact.
%
%
\subsection{Percentage of conjunction errors}
Differently form  \cite{osherson}, the papers \cite{Kahn-Tversky-conj-fall,Yates} study the percentage of
agents which judge $P(a_1, b_1)$ greater than $P(b_1)$.
In \cite{Yates} it is found that the fraction of agents that
exhibit single conjunction errors is greater for HL situations (0.67) than for HH and LL (0.15 and 0.27 respectively).
On the contrary, double conjunction errors happen more frequently for HH (0.30).

Our formalism can explain these results
by decomposing the density matrix into
$\rho=\sum_{i}P'_i|s'_i\rangle + \sum_{j}P''_j|s''_j\rangle$
where the states $|s'_i\rangle$ manifest conjunction fallacy, while the others
do not manifest it. Thus the probability that an agent will exhibit a conjunction fallacy,
being prepared in the state $\rho$, is $\sum_i P'_i$.
This means that in HL situations the states involved in the mixed state evidence strongly negative  phase factor
with more probability than for HH and LL, consistently with the results for the
mean conjunction errors.

We now consider the occurrence of double conjunction errors, which means
that $P(b_1)<P(a_1 \rightarrow b_1)$ and $P(a_1)<P(b_1\rightarrow a_1)$.
Equation  (\ref{interference_b}) entails a condition of maximal conjunction error similar to equation
(\ref{condition1}), that is
\begin{equation}\label{condition2}
\Delta_{max}(B,A)=P(b_0)P(a_1|b_0)-2\sqrt{P(a_1)P(b_1)P(a_1|b_1)P(b_1|a_1)}<0
\end{equation}
This condition entails that the allowed couples of probability $P(a_1)/P(b_1)$ for which there is a not-null
maximal conjunction error is HH, LH and LL. Thus the conditions HL and LH  do not allow for
double conjunction errors.
This can explain the results presented in \cite{Myamoto, Yates}, where several
probability combination models for conjunction errors are presented: double conjunction errors
can not occur simultaneously for HL or LH situations (maximal percentage 4.5\%).
\subsection{Numerical example}
We consider again the example of Linda, by considering two questions $A$: is Linda
feminist? and $B$: is Linda a bankteller? for which the experiments indicate a
probability $P(b_1) = 0.241$ and $P(a_1)P(b_1|a_1)=0.439$. Moreover, we know that $P(a_1)$
is high, since the description of Linda is very representative of a feminist.
In the quantum formalism, there are many different states describing such
a situation.

If we consider the pure state $\sqrt{P(a_0)}|a_0\rangle +
e^{i\phi_a}\sqrt{P(a_1)}|a_1\rangle$, we can choose  the parameters
$P(a_1)$ and $\phi+\xi-\phi_a$ such that equation (\ref{interference}) is
valid. For example, let be $P(a_1)=0.9$, from which we have
$P(b_1|a_1)=0.48$: thus $P(a_0)P(b_1|a_0)=0.05$. Let us now evaluate the
interference term
$2\sqrt{P(a_0)P(b_1|a_0)P(a_1)P(b_1|a_1)}cos(\phi+\xi-\phi_a)$, which is $0.3
cos(\phi+\xi-\phi_a)$. Thus from equation (\ref{interference}) we have
$0.241=0.05+0.439+0.3 cos(\phi+\xi-\phi_a)$, which entails
$cos(\phi+\xi-\phi_a)=-0.827$. This confirms the prediction of our model,
which evidences the presence of conjunction errors for situations
HL and a strongly negative phase factor for incompatible
events.
In this case we have built a pure state, but we could write more complex mixed states,
manifesting the conjunction error only with a certain probability.
\section{Other fallacies}\label{section:new}
The rich mathematical formalism of quantum mechanics and the  interference
effects allow us to predict or explain other fallacies in the
bounded-rationality regime. For example, 1) the ordering effects, 2) the
disjunction effect, 3) the conditional probability fallacy, 4) the framing
effect and 5) the uncertainty effect. We give in this final section a
brief description of these effects, which will be described in other
articles.

1) In the bounded-rationality regime, the order with which we  consider
two questions $A$ and $B$ is important (see also the case of rational
ignorance \cite{Rfranco_rat_ign_1}). Similarly to the repeated
Stern-Gerlach experiment, we ask a question relevant to the operator
$\widehat{A}$, then a second question relevant to the non-commuting
operator $\widehat{B}$ and finally again $\widehat{A}$. Equation
(\ref{joint_cond2}) defines the probability that the result of the second
question is $b_j$, given $a_i$. But the third measure leads to a non-null
probability $P(a_k \rightarrow b_j)$ for $k\neq i$: in other words the
third question can give a result for the observable $\widehat{A}$
different from the first.  The bounded rationality situation has
manifested an irrational behavior of the agent. The question $\widehat{A}$
has been asked two times, but what has been changed is the
\textit{context}. The opinion state of the agent in the test has evidenced
a contextuality effect. Thus it has great importance, in bounded
rationality, also the temporal order of the different questions.

2) The disjunction effect is an intriguing phenomenon discovered  by
Tversky and  Shafir \cite{Shafir} with important consequences in modelling
the interactions between inference and decision. This effect, like the
conjunction fallacy, considers the probabilities relevant to two events
which can be associated to two non-commuting operators. A first attempt to
give an explanation of the effect within the quantum formalism has been
given by \cite{Busemeyer}, by considering the two questions relevant to
two different Hilbert spaces. The consequences of this approach are that
we obtain an entangled state, but also that the evolution of the initial
state can lead to a state which contradicts the initial information given
to the agent. A new explanation of the disjunction effect will be given in
a separate paper: here we only observe that the conjunction fallacy can be
applied to show how the perceived probability $P(b_1)$ (for example) can
be lower than $P(a_0)P(b_1|a_0)+P(a_1)P(b_1|a_1)$, because the
interference terms like in equation (\ref{interference}) may appear.

3) The framing effects may have a similar explanation of the conjunction
and disjunction effect: the interference terms in fact are able to lower
the probability $P(b_j)$ when the initial information lead to an opportune
pure state in a basis relevant to a non-commuting operator.

4) The conditional probability fallacy is the assumption that
$P(a_j|b_i)=P(b_i|a_j)$.  In classical probability theory this is not
valid in general, while in quantum mechanics it is always true, as can be
seen from equation (\ref{conditional_p}).

5) Finally we make the hypothesis of the existence of an uncertainty
effect, which is  a consequence of the well-known uncertainty principle in
quantum mechanics. The experimental data of \cite{osherson} show that a
not-null percentage of agents in the tests fails in the implication $X
-and- Y \models X,Y$. We argue that this percentage may change, depending
on the commutator of the associated operators $\widehat{X},\widehat{Y}$.
The uncertainty principle in Hilbert spaces with discrete dimension has
been formulated in a more general form \cite{lur}: given the uncertainty
of an observable $\widehat{X}$, defined as the statistical variance of the
randomly fluctuating measurement outcomes $ \delta^2(\widehat{A})=\langle
\widehat{A}^2 \rangle - \langle \widehat{A} \rangle ^2$, the local
uncertainty relations \cite{lur} state that, for any set
$\{\widehat{X},\widehat{Y},\widehat{Z},...\}$ of non-commuting operators,
there exist a non-trivial limit $U$ such that
$\delta^2(\widehat{X})+\delta^2(\widehat{Y})+\delta^2(\widehat{Z})+...
\geq U$. This new form of the uncertainty principle may apply to
bounded-rationality regime, leading to predictions similar to those cited
of \cite{osherson}: even if we try to prepare the opinion state of agents
such that the uncertainty of $X$ and $Y$ is null, the sum
$\delta^2(\widehat{X})+\delta^2(\widehat{Y})$ can not be null.

All these paradoxical effects in general are due to the usual belief that
we can  assign pre-defined elements of reality to individual observables
also in regime of bounded rationality. In a classical situation, if we ask
to an agent the two questions associated to the observables $\widehat{A}$,
$\widehat{B}$, we can consider simultaneously the two answers $(a_i,b_j)$,
and we can study the joint probability $P(a_i,b_j)$. In a bounded
rationality regime, instead, this is not possible if the related
observables are non-commuting. We say that the answers to these questions
can not be known contemporarily, thus giving an important limit to the
complete knowledge of the opinion state of an agent in bounded-rationality
regime.

This effect in microscopic world is called quantum contextuality
\cite{contextuality},  and evidences, for any measurement,  the influence
of other non-commuting observables previously considered.
%

%
%
\section{Conclusions}\label{conclusions}
This article, addressed both to quantum physicists and to experts of
cognitive science,  evidences the incompatibility of quantum formalism
with Bayes' rule of classic probability theory, by deriving the violation
of equation (\ref{Conj_fallacy_classic})  in bounded-rationality regime.
In particular, we use mathematical objects like vector state and density
matrix to describe the opinion state of agents, and hermitian operators
for the questions: in section \ref{section:conjunction} we show that the
conjunction fallacy can be explained as an interference effect when two
different questions (relevant to two non-commuting operators) are
considered.

This seems to confirm the comment of \cite{osherson}, for which the
conjunction  fallacy seems to involve failure to coordinate the logical
structure of events with first impressions about chance. The first
impressions about chance may be encoded in the quantum phase, which leads
to interference effects. In fact, we have seen that state (\ref{S-A})
do not differ for the statistical predictions of $A$, but
for the presence of a phase, which gives us the information of how the
same superposition of states is considered initially by the agent.

Thus we conclude that the conjunction fallacy can be considered as a
natural  consequence of the quantum formalism used to describe the
bounded-rationality regime. By the way, the formalism introduced does not
only give an explanation of the fallacy, but also has a predictive
character: in fact, we have predicted a reverse conjunction fallacy, for
which the probability of the conjunction of two events is much less than
the probability assigned to a single event. Moreover, in section
\ref{section:new} we propose other effects which are consequences of the
formalism introduced.
\\\\\\
I wish to thank Jerome Busemeyer, Peter Bruza and Daniel Osherson for
making very useful comments and suggestions.
 \footnotesize
%
%


\section*{References}

\begin{thebibliography}{99}
%
\bibitem{economy}
Schaden, M. (2002). Quantum finance: A quantum approach to stock price fluctuations.
Physica A, 316, 511.

Baaquie, B. E. (2004). Quantum Finance: Path Integrals and Hamiltonians for Options
and Interest Rates. Cambridge UK: Cambridge University Press.

Haven, E. (2005). Pilot-wave theory and financial option pricing. International Journal
of Theoretical Physics, 44, 1957-1962.

Bagarello, F. (2006). An operatorial approach to stock markets. Journal of Physics A,
39, 6823-6840.
%
%
%
\bibitem{management}
Bordley, R. F. (1998). Quantum mechanical and human violations of compound probability principles:
Toward a generalized Heisenberg uncertainty principle. Operations
Research, 46, 923-926.

Bordley, R. F., \& Kadane, J. B. (1999). Experiment dependent priors in
psychology and physics. Theory and Decision, 47, 213-227.

Mogiliansky, A. L., Zamir, S., \& Zwirn, H. (2006). Type indeterminacy: A
model of the KT(Kahneman-Tversky)-man [preprint].
http://lanl.arxiv.org/abs/physics/0604166
%
\bibitem{Rfranco_rat_ign_1} R. Franco, Quantum mechanics and rational ignorance, arXiv:physics/0702163
%
%
\bibitem{psychology}
Grossberg, S. (2000). The complementary brain: Unifying brain dynamics and modularity.
Trends in Cognitive Science, 4, 233-246.

Gabora, L., \& Aerts, D. (2002b). Contextualizing concepts using a
mathematical generalization of the quantum formalism. Journal of
Experimental and Theoretical Artificial Intelligence, 14, 327-358.
Preprint at http://arXiv.org/abs/quant-ph/0205161

Busemeyer, J. R., Wang, Z., \& Townsend, J. T. (2006). Quantum dynamics of
human decision making. Journal of Mathematical Psychology, 50, 220-241.

\bibitem{aerts3}
Aerts, D. (2007a). A quantum field model for the conjunction of concepts. Manuscript submitted for publication.
Aerts, D. (2007b). A fundamental model for general concept formation. Manuscript submitted for publication.
%
%
%
\bibitem{game}

Eisert, J., Wilkens, M., \& Lewenstein, M. (1999). Quantum games and
quantum strategies. Physical Review Letters, 83, 3077-3080.

Piotrowski, E. W., \& Sladkowski, J. (2003). An invitation to quantum game theory.
International Journal of Theoretical Physics, 42, 1089.

%
%
%
\bibitem{language}

Widdows, D. (2003). Orthogonal negation in vector spaces for modelling word-meanings
and document retrieval. In Proceedings of the 41st Annual Meeting of the Association for
Computational Linguistics (pp. 136-143). Sapporo, Japan, July 7-12.

Widdows, D., \& Peters, S. (2003). Word vectors and quantum logic: Experiments with
negation and disjunction. In Mathematics of Language 8 (pp. 141-154). Indiana:
Bloomington.

Aerts, D., \& Czachor, M. (2004). Quantum aspects of semantic analysis and symbolic
artificial intelligence. Journal of Physics A, Mathematical and Theoretical, 37, L123-L132.

Bruza, P. D. and Cole, R. J. (2005). Quantum logic of semantic space: An exploratory
investigation of context effects in practical reasoning. In S. Artemov, H. Barringer,
A. S. d'Avila Garcez, L.C. Lamb, J. Woods (Eds.) We Will Show Them: Essays in
Honour of Dov Gabbay. College Publications.
%
%
%
%
\bibitem{Kahn-Tversky-conj-fall}  A. Tversky and D. Kahneman, (1983). Extension versus intuititve reasoning:
The conjunction fallacy in probability judgment. Psychological Review, 90, 293–315.\\
Tversky, A. and Kahneman, D. (1982). Judgments of and by
representativeness. In D. Kahneman, P. Slovic and A. Tversky
(Eds.), Judgment under uncertainty: Heuristics and biases.
Cambridge, UK: Cambridge University Press.
\bibitem{osherson}
E. Shafir, E. E. Smith, and D. Osherson, Typicality and reasoning fallacies.
Memory and Cognition, 18(3):229-239, 1990

Tentori K., Bonini N., Osherson D., "The conjunction fallacy: a misunderstanding about conjunction?".
Cognitive science : a multidisciplinary journal incorporating artificial intelligence, linguistics, neuroscience,
philosophy, psycology, 2004, v. 28, n. 3, p. 467-477,
\bibitem{bounded} H. Simon (1957). A Behavioral Model of Rational Choice, Models of Man, Social and Rational:
Mathematical Essays on Rational Human Behavior in a Social
Setting. New York: Wiley.
%
\bibitem{khrennokov} E. Conte, O. Todarello, A. Federici, F. Vitiello, M. Lopane, A. Khrennikov,
A Preliminar Evidence of Quantum Like Behavior in Measurements of Mental States,
arXiv:quant-ph/0307201v1
%
%
\bibitem{Rfranco_conj1} R. Franco, Quantum mechanics, Bayes' theorem and the conjunction fallacy, arXiv:quant-ph/0703222v1

\bibitem{bordley} Bordley, R. F. (1998) Quantum mechanical and human violations of compound probability principles.
Operations Research, 46, 923-926.

\bibitem{Dirac} P.A.M Dirac, Principles of Quantum Mechanics, 1930
%
\bibitem{contextuality}
A. Peres. Incompatible results of quantum measurements. Phys.
Lett. A, 151(3-4):107–108, 1990.

N. D. Mermin. Simple unified form for the major
no-hidden-variables theorems. Phys. Rev. Lett., 65(27):3373–3376,
December 1990.
%
\bibitem{Rfranco_rat_ign_3} R. Franco, The inverse fallacy and quantum formalism
http://xxx.lanl.gov/abs/0708.2972v1

\bibitem{Shafir} Shafir, E. and Tversky, A. (1992) Thinking through
uncertainty: nonconsequential reasoning and choice. Cognitive Psychology, 24, 449-474.
%
\bibitem{confirmation} K. Tentori, V. Crupi, and D. Osherson (2006).
Determinants of Confirmation. Psychonomic Bulletin \& Review, in press.

\bibitem{Myamoto} Miyamoto, J. M., Gonzalez, R. and Tu, S. (1995). Compositional anomalies in the semantics of evidence. In J.
Busemeyer, R. Hastie and D. Medin (Eds.), Decision making from a cognitive perspective. (Volume 32 of the Psychology of
Learning and Motivation). New York: Academic Press.
%
\bibitem{Yates} Yates, J. F. and Carlson, B. W. (1986). Conjunction errors: Evidence for multiple judgment procedures, including
"signed summation." Organizational Behavior and Human Decision Processes, 37, 230-253.
%
\bibitem{Hampton} Hampton, J. A. (1988b). Disjunction of natural concepts. Memory \& Cognition, 16, 579-591.
%
\bibitem{Busemeyer}
Busemeyer, J. R., Matthew, M., and Wang, Z. (2006) An information processing explanation of disjunction effects. In R.
Sun and N. Miyake (Eds.) The 28th Annual Conference of the Cognitive Science Society and the 5th International
Conference of Cognitive Science (pp. 131- 135). Mahwah, NJ: Erlbaum.

%
%
\bibitem{lur} Hofmann H F and Takeuchi S 2003 Phys. Rev. A 68
032103,

Guhne O 2004 Phys. Rev. Lett. 92 11
%

%
\end{thebibliography}
\end{document}